# Experiment Neutrino-4 and restrictions for sterile neutrino

A.P. Serebrov[1], V.G. Ivochkin[1], R.M. Samoilov[1], A.K. Fomin[1], A.O. Polyushkin[1], V.G. Zinoviev[1],
P.V. Neustroev[1], V.L. Golovtsov[1], A.V. Chernyj[1], O.M. Zherebtsov[1], M.E. Chaikovskii[1], V.P. Martemyanov[2],
V.G. Tarasenkov[2], V.I. Aleshin[2], A.L. Petelin[3], A.L. Izhutov[3], A.A. Tuzov[3], S.A. Sazontov[3], M.O. Gromov[3],
V.V. Afanasiev[3], M.E. Zaytsev[1, 4], D.K. Ryazanov[4]

1. *NRC "KI" Petersburg Nuclear Physics Institute, Gatchina,*
2. *NRC "Kurchatov Institute", Moscow,*
3. *JSC "SSC Research Institute of Atomic Reactors", Dimitrovgrad, Russia,*
4. *Dimitrovgrad Engineering and Technological Institute MEPhI, Dimitrovgrad, Russia*

## Abstract

The experiment "Neutrino-4" started in 2014 on a model, then it was continued on a full-scale detector, and now, has provided the measurement result on dependence of the flux and spectrum of reactor antineutrinos on the distance of 6 -12 meters from the center of the reactor. One of the main problems is the correlated background from fast neutrons caused by space radiation. Attempts to suppress the background of fast neutrons by sectioning the detector have given some result. The relation of signal/background has improved up to 0.6. As a result, measurements of the difference in the counting rate of neutrino-like events (reactor ON–reactor OFF) have been obtained as dependence on distance from the reactor center. The fit of experimental dependence with the law $1/L^2$ give satisfactory result. The goodness of that fit is 81%. However, there was discovered experimental neutrino spectrum difference from calculated one. With achieved accuracy this difference does not change with distance. Therefore it cannot be interpreted as oscillations. Calculated spectrum form correction for experimental allow us to make proper analysis of oscillation parameters $\Delta m_{14}^2$ and $\sin^2(2\theta_{14})$ limitations. Result of this analysis is exclusion of reactor and gallium anomalies area with 95% CL. Experiment future perspectives are discussed.

## 1. INTRODUCTION

At present, there is a widely spread discussion on possible existence of a sterile neutrino having much less cross-section of interaction with matter than, for example, a reactor electron antineutrino. It is assumed, that owing to a reactor antineutrino transition to a sterile state, the oscillation effect at a short reactor distance and deficiency of a reactor antineutrino flux at a long range are likely to be observed [1, 2]. Moreover, a sterile neutrino can be considered as a candidate for the dark matter.

Ratio of the neutrino flux observed in experiments to the predicted one is estimated as 0.934 ± 0.024 [3]. The effect comprises 3 standard deviations. This is not yet sufficient to have confidence in existence of the reactor antineutrino anomaly. The method for comparing the measured antineutrino flux with the expected one from the reactor is not satisfactory, because of the problem of accurate estimating a reactor antineutrino flux and efficiency of an antineutrino detector. In Fig. 1 is shown a possible process of oscillations to a sterile state at small distances from an active zone of the reactor, which is presented in work [1].

The idea of oscillation can be verified by direct measuring of a flux distance dependence and antineutrino spectrum at a short reactor distance of 6 – 12m. A detector is supposed to be movable and a spectrum sensitive. Our experiment focuses on the task of confirming possible existence of a sterile neutrino at a certain confidence level or disproving it. For searching oscillations to a sterile neutrino, it is necessary to register discrepancy of the reactor antineutrino flux distance dependence from $1/L^2$. For the experiment to be conducted, one needs to make measurements of the antineutrino flux and spectrum at short distances from, practically, a point source of antineutrino.

If such a process does occur, it can be described by an oscillation equation:

$$P(\bar{\nu}_e \to \bar{\nu}_e) = 1 - \sin^2(2\theta_{14})\sin^2(1.27\frac{\Delta m_{14}^2 L}{E_{\bar{\nu}}}) \quad (1),$$

where $E_{\bar{\nu}}$ is antineutrino energy, with oscillations parameters $\Delta m_{14}^2$ and $\sin^2(2\theta_{14})$ being unknown.

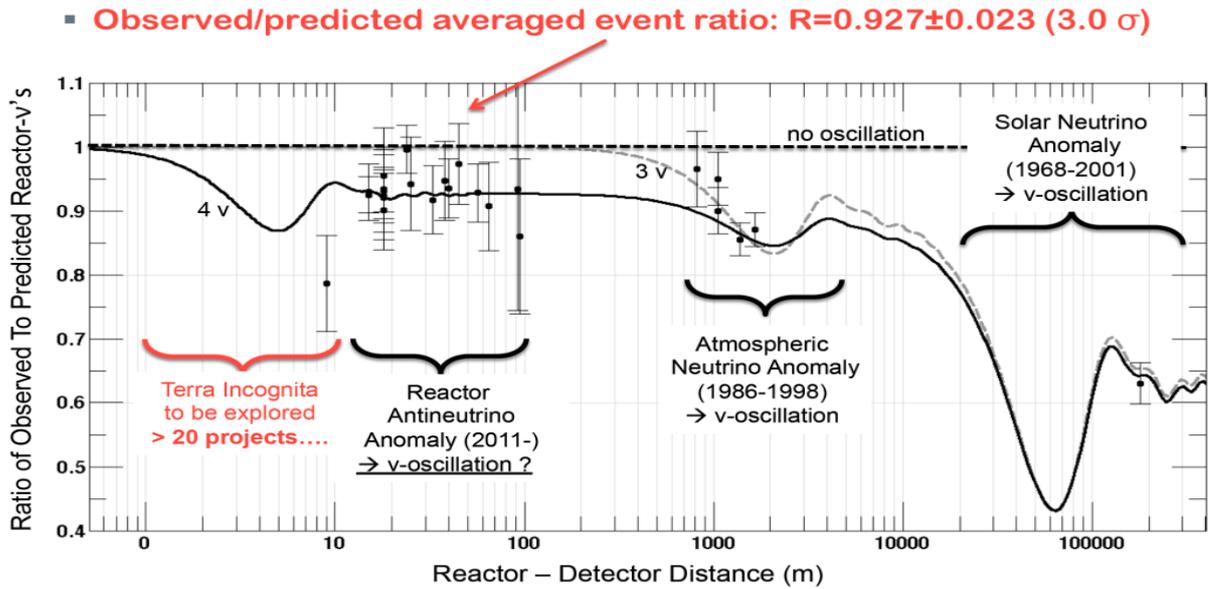

FIG.1. The possible process of oscillations to a sterile state at small distances from an active zone of the reactor presented in [1].

We have studied opportunities for making new experiments at research reactors in Russia. The research reactors should be employed for performing such experiments, since they possess a compact reactor core, so that a neutrino detector can be placed at a sufficiently small distance from it. Unfortunately, the research reactor beam hall has quite a large background of neutrons and gamma quanta, which makes it difficult to perform low background experiments. Due to some peculiar characteristics of its construction, reactor SM-3 provides the most favorable conditions for conducting an experiment on search for neutrino oscillations at short distances [4,5]. But, SM-3 reactor is located on the Earth surface, hence, cosmic background is the major difficulty in making such experiments by research reactors.

**2. REACTOR SM-3**

Initially, 100-megawatt reactor SM-3 was designed for carrying out both beam and loop experiments. Five beam halls were built, separated from each other with big concrete walls as wide as ~1 m (Fig. 2). This enabled to carry out experiments on neutron beams, without changing background conditions at neighboring installations. Later on, the main experimental program was focused on the tasks concerned with irradiation in the reactor core center. For 25 years of exploitation, a sufficiently high fluence was accumulated on materials of the reactor cover, which necessitated its replacement. Setting a new reactor cover on new reactor tank was the simplest way to solve this problem. Such a solution, however, resulted in raising a reactor core center by 67 cm higher than in the previous position.

Horizontal beam channels were removed, as priority was given to conducting loop experiments. Neutron flux in the location of the former beam channels was lowered by four orders of magnitude. Respectively, it caused decrease in the neutron background to about $4 \cdot 10^{-3}$ n/cm$^2$s (on thermal neutrons) in the former beam halls. It is approximately by 4 – 5 orders of magnitude lower than a typical neutron background in the beam hall of a research reactor. Lately, in making preparations for an experiment on search for oscillations of a reactor antineutrino to a sterile state at reactor SM-3, upgrading of a slide valve of the former neutron beam has been completed. As a result, the background of fast neutrons has diminished to the level of $10^{-3}$ n/cm$^2$s, i.e. practically, to the level of neutron background on the Earth surface, caused by space radiation. These conditions are most preferable of all possible for a neutrino experiment to be performed. Other advantages of SM-3 reactor are a compact reactor core center (35×42×42 cm), with high reactor power equal to 100 MW, as well as a sufficiently short distance (5 m) from the center of a reactor core to the walls of an experimental hall. Besides, of special significance is the fact that antineutrino flux can be measured within a sufficiently wide range from 6 to 12 meters.

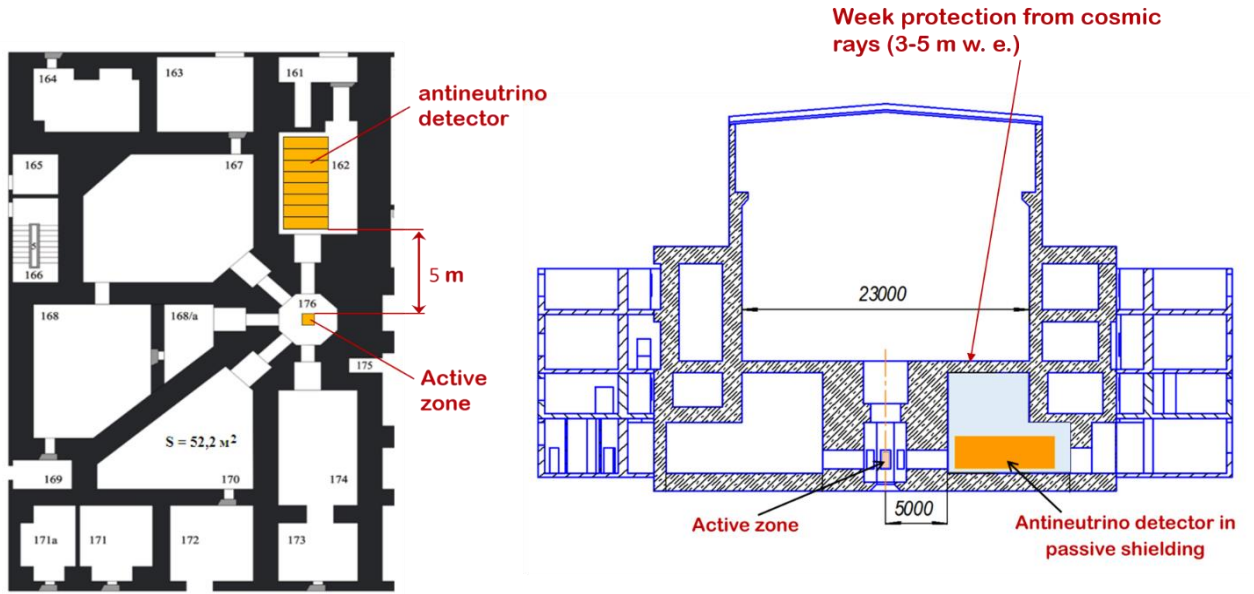

FIG.2. Detector location at reactor SM-3.

## 3. PASSIVE SHIELDING OF A NEUTRINO DETECTOR AT SM-3 REACTOR

Layout of passive shielding ("cabin") from the outside and inside is given in Fig. 3. It is created from elements based on steel plates of 1x2 m, 10mm thick, to which are attached 6 lead sheets of 10 mm thickness. The cabin volume is 2x2x8 m. From the inside, the cabin is covered with plates of borated polyethylene of 16 cm thickness. The total weight of passive shielding is 60 tons, the volume of borated polyethylene is $10m^3$. Inside the passive shielding, there is a platform with an antineutrino detector, which can be moved along the rails within the range 6 - 12 meters from the reactor core center. A neutrino channel can be entered by means of a ladder, through the roof with the removed upper unit, as shown in Fig. 3. Loading of the detector into a neutrino channel is carried out from the main hall, through a trap door in the building ceiling. In this case, an overhead crane of the main hall is used.

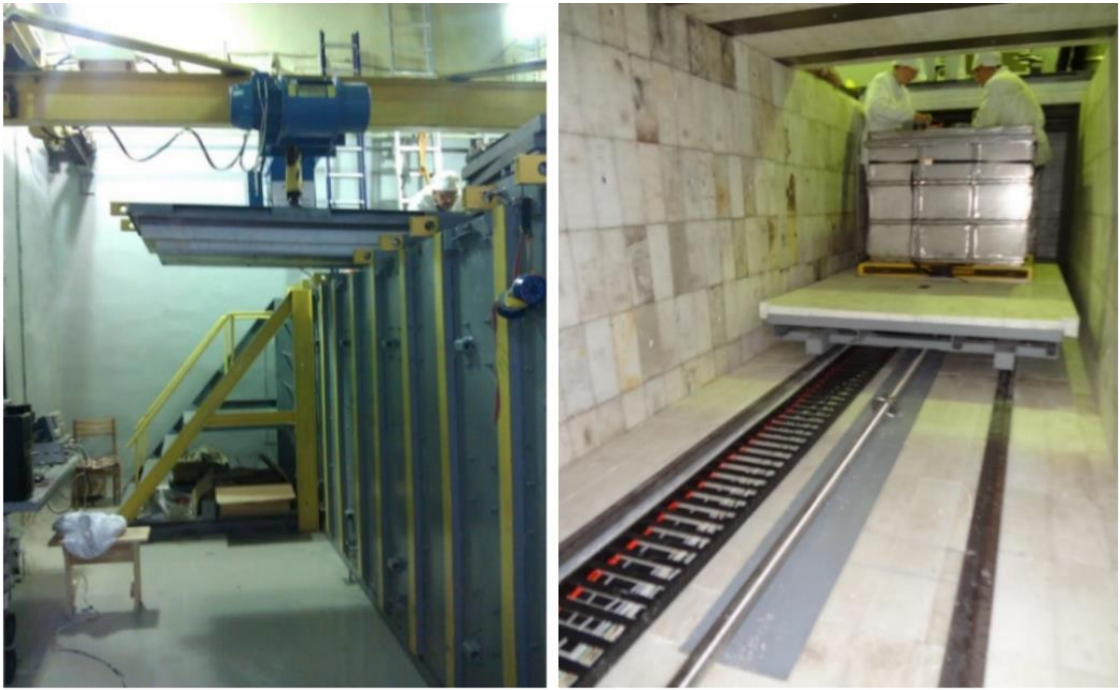

FIG.3. General view of passive shielding: from the outside and inside. The range of detector dislocation is 6 - 12 m from the reactor core center.

## 4. INVESTIGATION OF BACKGROUND CONDITIONS WITH A GAMMA DETECTOR

For measuring the gamma ray spectrum, a detector NaJ(Tl) of 60×400 mm was applied. In switching on the reactor, it shows counts of gamma-quanta from a neutron capture in an iron-concrete shielding of the reactor. During a reactor operation, the background of gamma radiation, in the energy range from 3 MeV to 8 MeV significantly increase (22 times larger than reactor Off state), because of thermal neutrons interaction with iron nuclei in construction materials. This energy range is of great importance, since it corresponds to gamma-quanta energy at the neutron capture by Gd.

Measurements of gamma-rays background inside passive shielding results are shown in Fig. 4. Gamma-radiation of isotopes $^{137}Cs$, $^{60}Co$ is independent of the reactor operation mode and is caused by radioactive contamination from the building floor and walls. In spite of pouring of the concrete floor with iron grit and reconstruction of a slide valve, which reduced 5 – 6 times gamma radiation background in this energy range, it does remain high enough, which confirms necessity of passive shielding from gamma quanta for a detector.

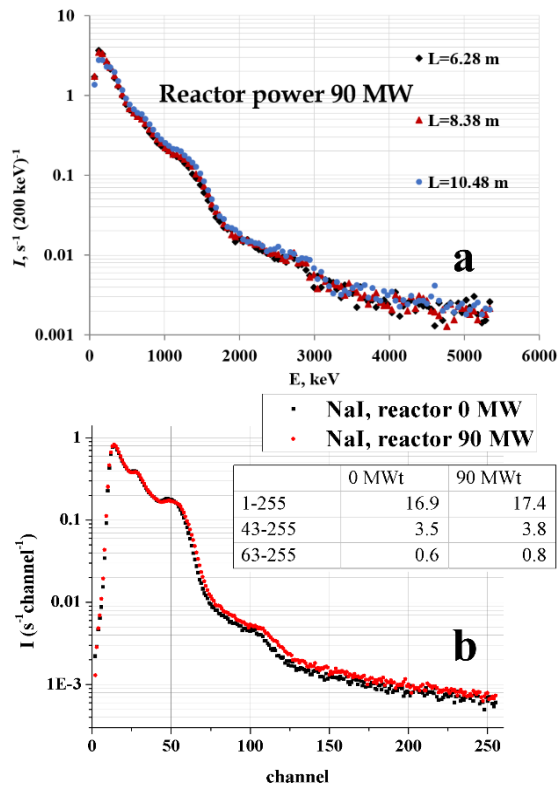

FIG.4. Gamma-radiation spectra at the detector location. a) Reactor power is 90MW. L is distance from the reactor core center: - 6.28 m, - 8.38m, - 10.48m. b) Reactor ON/OFF spectra.

Within energy range of 1440÷7200 keV (from $^{40}K$ and higher), the 5 cm lead shielding makes the level of background gamma radiation 4.5 times lower, which proves that its installation on the detector is reasonable. However, it is to be noted, that neutron background, resulting from the interaction of cosmic muons with lead nuclei, enhances inside the lead shielding. Indeed, the 5 cm lead shielding around a neutron detector results in raising twice its count rate. Hence, inside the lead shielding, there must be located another one, made from borated polyethylene.

Fig. 4a presents the gamma spectrum shape inside the passive shielding for different distances along the way of a neutrino detector: 6.28 m, 8.38 m, 10.48 m. No noticeable alterations in the spectrum shape is observed. Moreover, for comparison, gamma-spectra are measured at the reactor On and Off inside the passive shielding, at the point nearest to the reactor. Considerable difference in spectra is not found

## 5. ESTIMATIONS OF FAST AND THERMAL NEUTRON FLUXES

In 2013, at SM-3 reactor, a neutrino laboratory building was completed for exploitation and the equipment for passive shielding of a neutrino detector was mounted. The slide valve for the former neutron channel was carefully plugged. As a result, a flux of thermal neutrons in the neutrino laboratory building decreased 29 times to the level of $1 \div 2 \cdot 10^{-4} n/cm^2 s$.

This level is determined by space radiation neutrons and, practically, is independent of the reactor operation. Measurements of thermal neutron fluxes were made with $^3He$ detector, which is a proportional counter of 1m long with diameter of 30 mm. For registration of fast neutrons, we used the same proportional $^3He$ detector, but it was put into the shielding made of polyethylene (thickness of layer is 5 cm), which in its turn was wrapped in a layer of borated rubber (3 mm thick, containing 50% of boron). To convert count rate ($s^{-1}$) of proportional $^3He$ detector into density units of a prompt neutron flux ($cm^2 \cdot s^{-1}$), a newly created fast neutron detector was calibrated according to recordings of standard MKC AT6102. For this purpose both detectors were placed side by side at distance of 3 m from a neutron source (Pu-Be). Thermal neutrons detector ($^3He$ detector) was calibrated in the same way. Made in this way, detectors of thermal and fast neutrons have sensitivity nearly by two orders of magnitude higher than that of standard devices. They were employed for conducting low-background measurements by a neutrino laboratory. Estimations on neutron background were made, at first, before gate upgrading of the former neutron beam (before slide valve was plugged), then, after upgrading and, finally, after installing passive shielding for a neutrino detector. A detector of fast neutrons was located on the roof of passive shielding and directly on the reactor wall, i.e. at distance of 5.1 m from the reactor core center. Measurements results are in Fig.5.

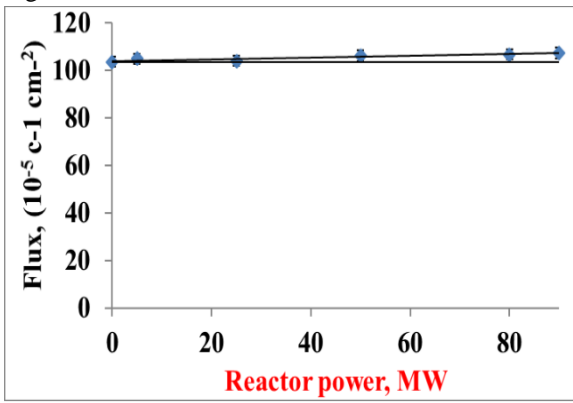

FIG.5. Flux of a fast neutron detector at the output of a power reactor 90MW. The counter is located on the cabin roof at the reactor wall.

From Fig. 5, one can conclude that background of fast neutrons, practically, does not depend on the reactor power. It is determined by neutrons emerging from interaction of muons with nuclei of the surrounding materials, in particular, with lead nuclei of passive shielding. However, suppression of the flux with internal lining of passive shielding (16 cm of borated polyethylene) is 12 times higher.

Another problem of great significance is reactor influence on the flux of fast neutrons inside passive shielding. Measurements of the fast neutron flux inside passive shielding at reactor On were in progress, at the nearest to the reactor wall, for about ten days, and for the same time after switching it off. At the reactor On the fast neutron flux was equal to $(5.54 \pm 0.13) \cdot 10^{-5} \text{s}^{-1} \text{cm}^{-2}$, while at the reactor Off, it was $(5.38 \pm 0.13) \cdot 10^{-5} \text{s}^{-1} \text{cm}^{-2}$, i.e. there was no difference within the accuracy of 2.5%.

Much more detailed measurements were made with a fast neutron detector on the cover of a neutrino detector, moved during these measurements along the neutrino channel in the range from 6.25 m to 10.5 m. The fast neutron detector was set on the cover of the neutrino detector and dislocated together with it. Results of these measurements at the reactor On and the reactor Off are given in Fig. 6.

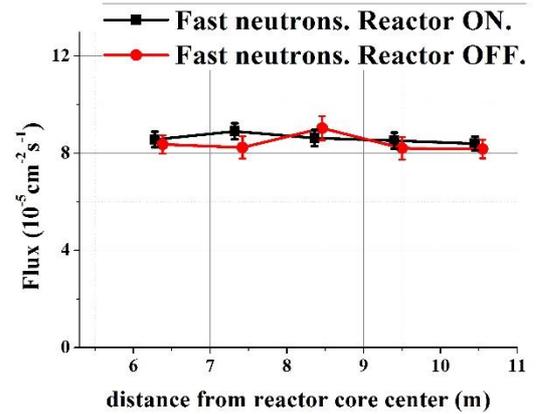

FIG.6. Fast neutron background at different distances from the reactor core center measured with the detector of fast neutrons inside passive shielding.

There is no difference in results at the reactor On and the reactor Off within statistical estimation accuracy. The background involved is determined by space radiation. In these measurements, the background level appeared to be equal to $(8.5 \pm 0.1) \cdot 10^{-5} \text{s}^{-1} \text{cm}^{-2}$, which is somewhat higher than that at the reactor wall. The discrepancy can result from the detector location regarding the direction of a neutron flux, i.e. its vertical position at the reactor wall as well as a horizontal one, on the cover of the neutrino detector.

## 6. FULL-SCALE ANTINEUTRINO DETECTOR

Detector scheme with active and passive shielding is presented at Fig. 7. The full-scale detector with a liquid scintillator has volume of 1.8 m³ (5x10 sections 0.225x0.225x0.85м³, filled to 70 cm). Scintillator with gadolinium concentration 0.1% is using to detect inverse beta decay (IBD) events $\tilde{v}_e + p \rightarrow e^+ + n$. The neutrino detector active shielding consists of external ("umbrella") and internal parts regarding passive shielding. The internal active shielding is located on the top of the detector and under it. First and last row of the detector are using as an active shielding and at the same time also as a passive shielding from fast neutrons. Thus, fiducial volume of scintillator is 1.42 m³. Detector has a sectional structure. For making measurements the detector can be moved to different positions on distance divisible by section size. As a result, different sections can be placed at the same places with respect to the reactor excepting edge effects at closest and the most distant positions.

Creation of a multi section system is aimed at finding criteria for detection of neutrino events. The main problem of an experiment on the Earth's surface is fast neutrons from space radiation. The scattering of fast neutrons easily imitates an IBD. Registration of the first (start or prompt) signal from recoil protons imitates registration of a positron. The second (stop or delayed) signal arises in both cases when a neutron is captured by gadolinium. The difference between these prompt signals is in appearance of two gamma quanta 511 keV each from annihilation of a positron from IBD process. Positron range in an organic scintillator is less than ~5cm. The recoil proton track with high probability is within the size of one detector section, because length of a recoil proton track is about ~1 mm. Gamma quanta with energy of 511 keV flying away in opposite directions can be registered in adjacent sections. Monte Carlo calculations shown that 63% of starting signals from neutrino events can be also recorded within one section. Only 37% of neutrino starting signals is multi-sectional due to registration of γ quanta in the sections adjacent to that, in which positron annihilation occurs [6]. Thus, if signal difference between the reactor ON and reactor OFF measurements is within 37% - 63% for the multi section and single-section events, then it can be interpreted as a neutrino signal.

In our measurements the difference in parts of count rates at the reactor On and Off for double and single prompt events, integrated over all distances is $(37 \pm 4)\%$ and $(63 \pm 7)\%$. This ratio allows us to treat the recorded events as neutrino events.

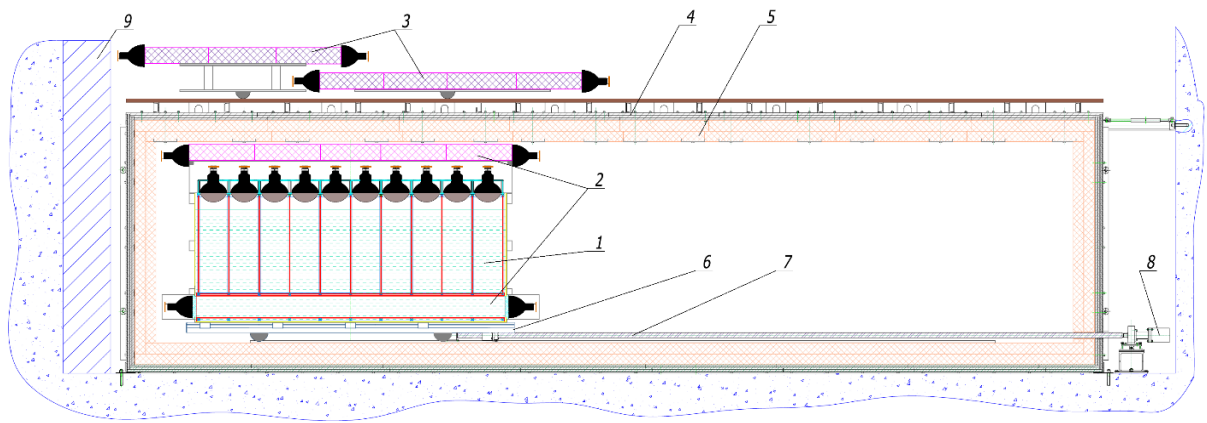

FIG. 7. General scheme of an experimental setup. 1 – detector of reactor antineutrino, 2 – internal active shielding, 3 – external active shielding (umbrella), 4 – steel and lead passive shielding, 5 – borated polyethylene passive shielding, 6 – moveable platform, 7 – feed screw, 8 – step motor, 9 –shielding against fast neutrons from iron shot.

### 7. ENERGY CALIBRATION, MEASUREMENTS WITH FULL-SCALE DETECTOR PROTOTYPE

As pointed out in the introduction, along with observation of the spatial flux discrepancy, it is of significance to study a neutrino spectrum in various detector positions. This task requires conducting energy calibration of the detector. Preliminary investigation was made with full-scale detector section analogue, which scheme is shown in Fig. 8.

Volume with mirror walls and bottom filled with water up to 75cm. There is an 8 cm gap between PMT and water surface. Plastic scintillator 10x10x5cm$^3$ with gamma-source Na$^{22}$ attached can be moved up and down. Spectra obtained with different positions of scintillator are shown in Fig. 9. Dependence of 1274keV peak on source $^{22}$Na position presented at Fig. 10. From Fig. 9 and Fig. 10 it is clear that source's spectrum weekly changes with scintillator position. This signal amplitude smoothing is due to light reflecting from mirror bottom and reflecting part of light yield from water surface because of full internal reflection.

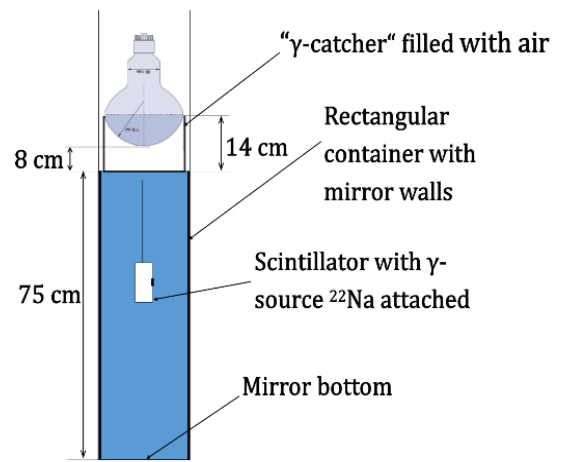

FIG. 8. Scheme of model to measure with full-scale detector section analogue.

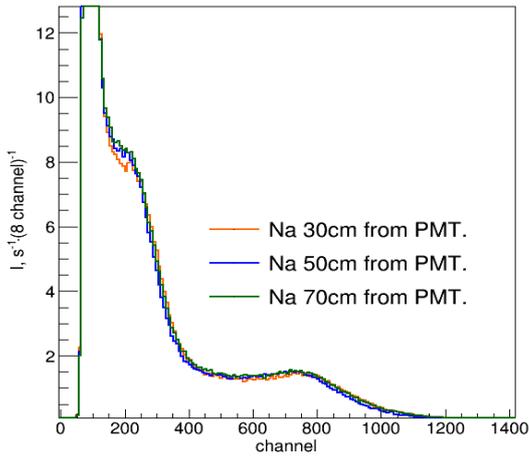

FIG. 9. $^{22}$Na source spectrum with different scintillator position for model of full-scale detector section with air gap "gamma-catcher".

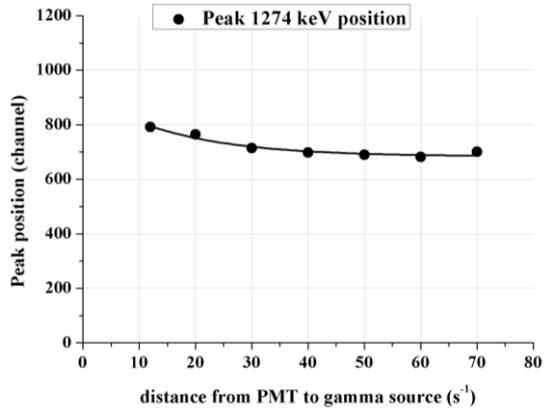

FIG. 10. Dependence of 1274 keV position from distance between $^{22}$Na source on plastic scintillator and PMT.

Before starting an experiment, a multi-section detector was calibrated using γ-quanta source ($^{22}$Na) and a neutron source (Pu-Be). The results of energy calibration are presented at Fig. 11 and Fig. 12. Calibration was made by placing γ-quanta source ($^{22}$Na) and neutron source (Pu-Be) above the detector, irradiating in this way, 2 rows of sections [7].

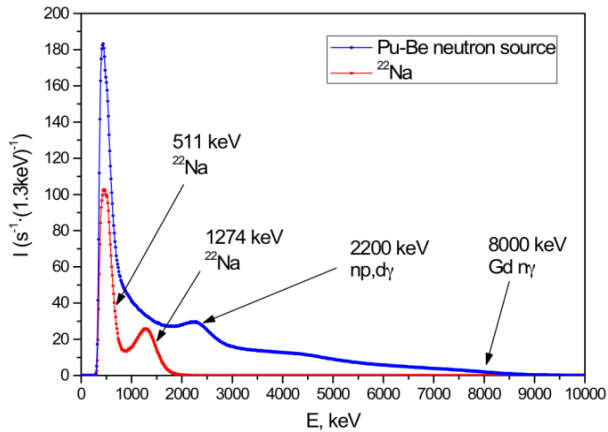

FIG. 11. Energy calibration of neutrino detector.

Calibration for 4.5MeV energy was made by fast neutron from Pu-Be source in additional experiment without moderators. Calibration results are shown in Fig. 12.

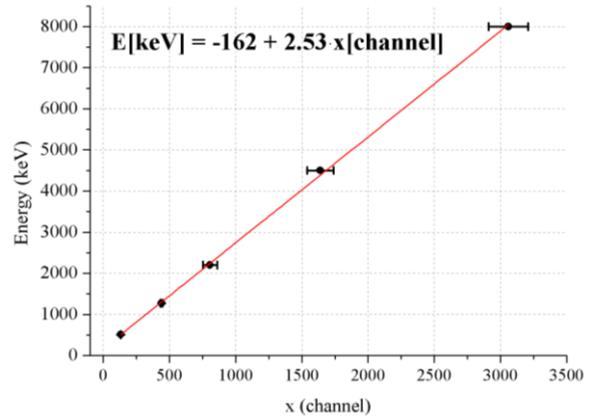

FIG.12. Results of energy calibration.

## 8. RESULTS OF DISTANCE DEPENDENCE MEASURMENTS

Measurements by a full-scale detector with started in June 2016. Measurements with the reactor ON were carried out for 429 days, and with the reactor OFF- for 219 days. In total, the reactor was switched on and off 51 times. First results of this work are presented.

Results of measurements on difference in counting rate of neutrino-like events for a full-scale detector are shown in Fig. 13, as dependence of antineutrino flux on the distance from the reactor center. A new analysis is made only with data on a full-scale detector after iron pellet filling the room between wall from reactor side and detector shielding. Therefore, the effect of fast neutrons from the reactor is completely excluded. Besides, the first and last rows of the detector are used as active protection, at the same time it can work as passive protection against fast neutrons.

Fit of an experimental dependence with the law $1/L^2$ yields satisfactory result. At the same time, there is an experimental result of DANSS, kindly provided by M. V. Danilov, and also presented at the 52nd Rencontres de Moriond [8], which points out that in the range of 10 – 13 meters the law $1/L^2$ holds true with rather good accuracy. We can make use of this information and combine results of measurements in the range of their overlapping of 10 – 12 meters. The goodness of the fit with combined data is 81%.

Using obtained experimental distance dependence we can do an analysis of limitation for oscillation parameters $\Delta m_{14}^2$ and $\sin^2(2\theta_{14})$. Results of that analysis presented in Fig. 14.

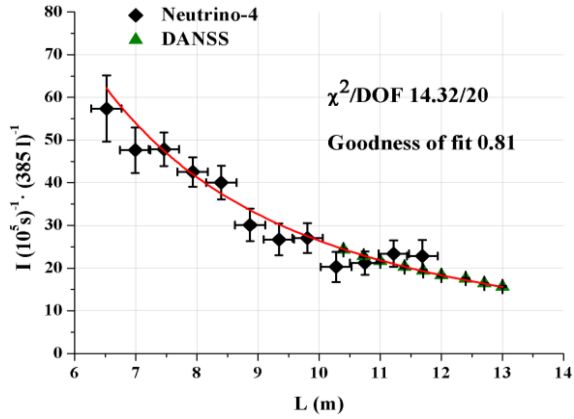

FIG. 13. Reactor antineutrino flux distance dependence for a full-scale detector, the point graph fit for dependence of $1/L^2$, where L – distance from the center of reactor core.

As can be noted at $\Delta m_{14}^2 \geq 1 eV^2$ this method is less sensitive to $\sin^2(2\theta_{14})$ value. It has to do with averaging of possible oscillation over wide spectrum of reactor antineutrino. To analyze that area spectral measurements needed.

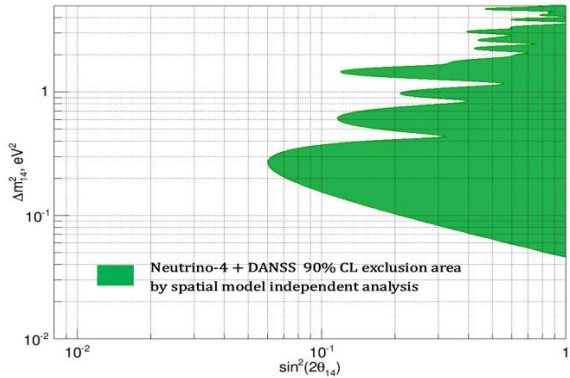

FIG. 14. Restrictions for oscillation to sterile state parameters followed only from the analysis of antineutrino flux dependence on distance to the reactor core center.

## 9. RESULTS OF SPECTRAL MEASURMENTS

The example of spectral measurements within two months at closest to the reactor position (distance between centers of the reactor and detector is 7.11 m) is presented in Fig. 15. The Fig. 16 illustrates the spectrum of prompt signals summarized over all distances for increasing statistical accuracy. This spectrum contains statistics of all measurements.

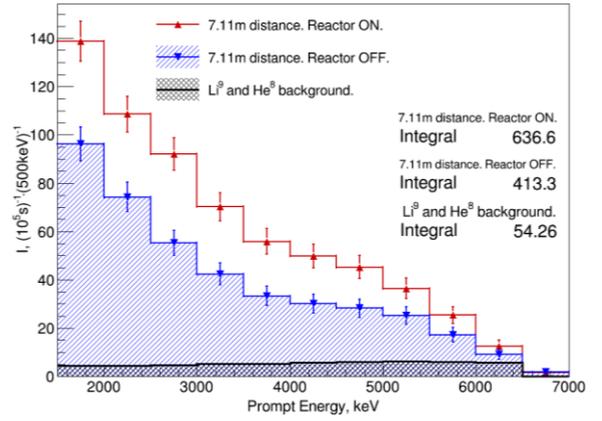

FIG. 15. Example of the spectrum of prompt signals obtained within two months of data taking. The signal (ON – OFF) has made 223 events per day. Relation signal/background (ON-OFF)/OFF = 0.57.

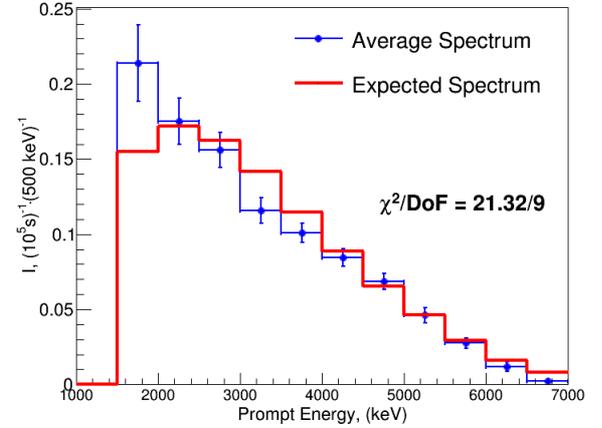

FIG. 16 Spectrum of prompt signals for a total cycle of measurements summarized on all distances (average distance - 8.6 meters). The red dotted line shows Monte -Carlo simulation with neutrino spectrum for $U^{235}$, as the SM-3 reactor works on highly enriched uranium.

There is a discrepancy between calculated and experimental spectra in 3 MeV region. Spectra were normalized to unity. The ratio of an experimental spectrum of prompt signals to the spectrum expected from MC calculations is presented in Fig. 17a for three distances 7.3 m, 9.3 m and 11.1 m. Averaged over all distances ratio and its polynomial approximation (red curve) are shown in Fig. 17b. It should be noted, deviation of experimental spectrum from calculated is identical for different distances within achieved accuracy. At least red curve is well fitted for every ratio and goodness of fit is 0.77, 0.78, 0.68 for 7.3м, 9.3м и 11.1м. respectively.

So-called "bump" in 5 MeV area also is observed just as in other experiments [9-13], but value of the deviation is larger than in experiments at atomic stations. If it is linked to $^{235}$U, assumed in [14-16], it could be explained by highly enriched fuel $^{235}$U (95%) at SM-3 reactor as distinct from effective fission fraction of $^{235}$U 56% [11] or 65% [9,10,13] at different industrial reactors.

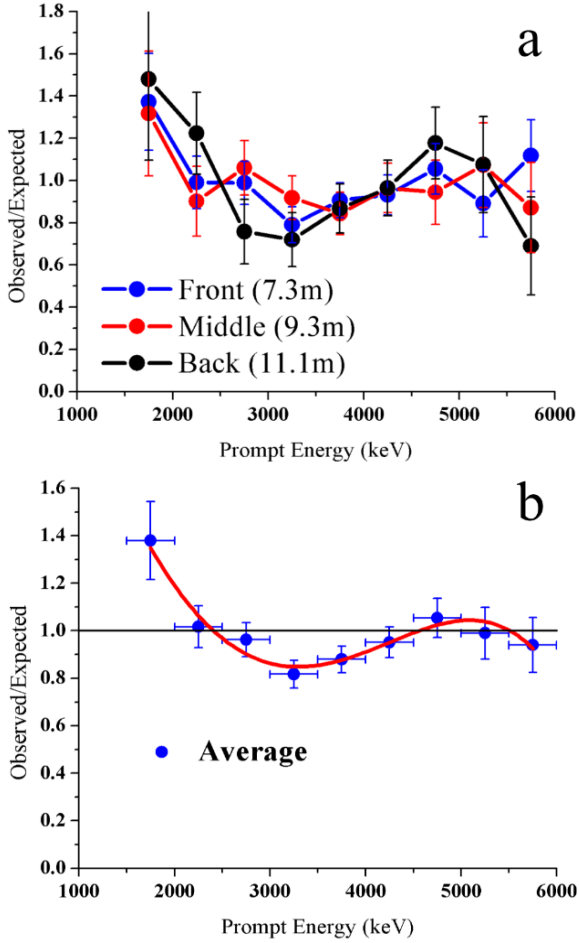

For the best fit parameters $\Delta m_{14}^2 = 0.60$ and $\sin^2(2\theta_{14}) = 0.28$ expected oscillation effect at different distances can be estimated.

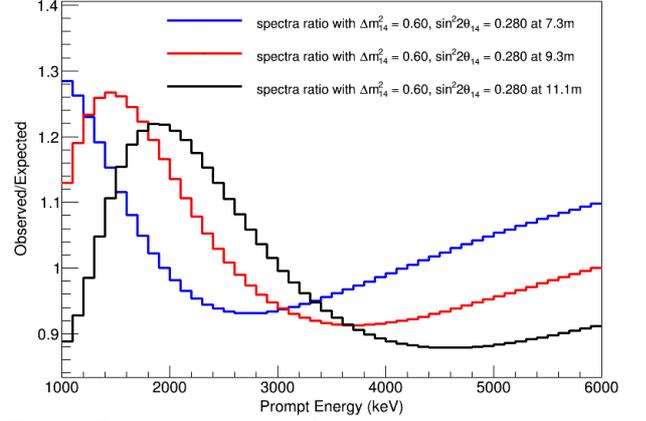

FIG. 19. Estimation of expected oscillation effect at different distances with parameters $\Delta m_{14}^2 = 0.60$ and $\sin^2(2\theta_{14}) = 0.28$

Comparison of Fig. 19 and 17a shows there is no distinctive energy shift with distance increasing. In particular, goodness of fit for experimental dependencies with curves from Fig. 19 is 31%, 89%, 7% for 7.3 m, 9.3 m and 11.1 m respectively. It should be noted expected oscillation parameters $\Delta m_{14}^2 = 0.6 \text{eV}^2$ and $\sin^2 2\theta_{14} = 0.28$ contradict to the good approximation of experimental flux distance dependence with $1/L^2$.

Thus, most probable cause of discrepancies is calculation. And, considering the deficiency of observed antineutrino flux compared with calculated 0.934 [3] not a "bump" should be discussed, but a "hole" around 3 MeV, which is, apparently, responsible for reactor antineutrino anomaly.

If we accept a hypothesis of calculations imperfections, we should make corrections to expected spectrum using dependence of deviation of calculated spectrum from experimental (red plot in Fig. 17). After that we can do an analysis for oscillations parameters once again. For that ratio of experimental spectra at different distances to calculated spectrum after correction could be presented as dependence from $L/E_{\bar{\nu}}$ in (1). Result of this analysis is shown in Fig. 20. Obtained dependence is very well fitted with constant equal one and $\chi^2$ is 11.98/27 and goodness of fit is 0.99.

FIG. 17. The ratio of an experimental spectrum of prompt signals to the spectrum, expected from MK calculations.

Assuming, discrepancies between experimental spectrum and calculated one are due to oscillation analysis for parameters $\Delta m_{14}^2$ and $\sin^2(2\theta_{14})$ limitations can be done. Results of such analysis is shown in Fig. 18.

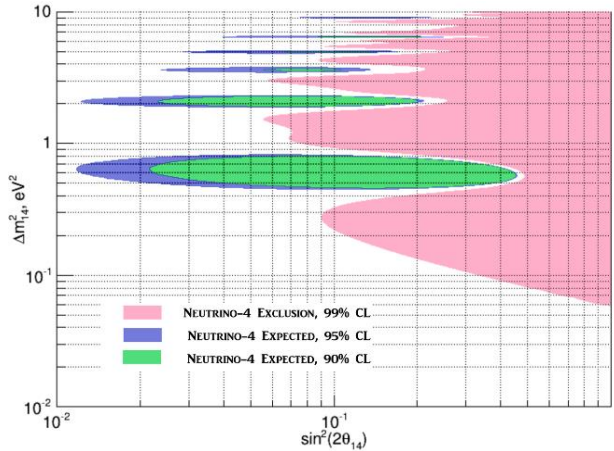

FIG. 18. Data analysis on the assumption that experimental spectrum distortion in comparison with calculated one is caused by neutrino oscillation to a sterile state. Red area excluded with 99% CL. Blue and green areas of expected parameters oscillation with 95% and 90% respectively.

Using these ratios of experimental spectra to calculated spectrum after correction as function of $L/E_{\bar{\nu}}$, new analysis for parameters of oscillation restrictions can be done. Result of that analysis is presented in Fig. 21.

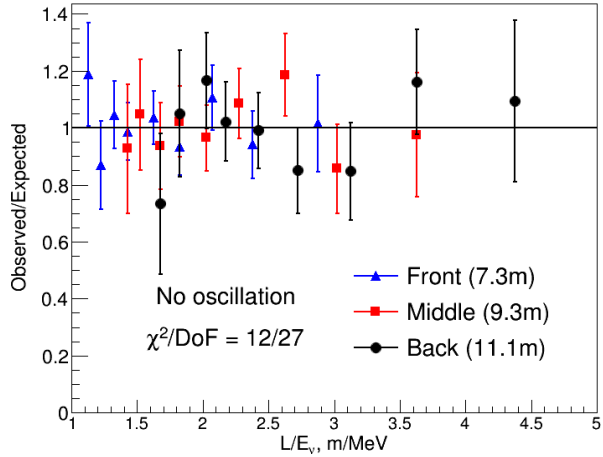

FIG. 20. Ratio of experimental spectra to calculated spectrum after correction as function of $L/E_{\bar{\nu}}$ for three distances: 7.3 m, 9.3 m and 11.1 m.

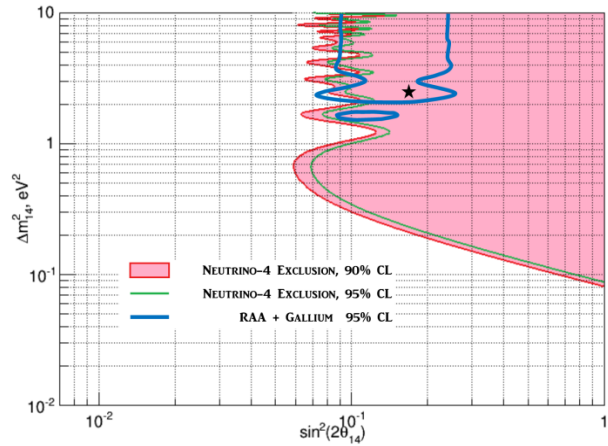

FIG. 21.. New analysis for oscillation parameters limitations after corrections of calculated spectrum.

From the analysis one can conclude that area of reactor and gallium anomalies in the framework of model with one sterile neutrino is excluded with 95% CL.

## 10. CONCLUSION ON THE CURRENT EXPERIMENTAL RESULTS

Now it is necessary to make conclusions from the carried-out analysis of experimental data.

The Neutrino-4 experiment started in 2014, at first on a model, then on a full-scale detector and, for the first time in the world, has provided measurement data on dependence of the reactor antineutrinos flux on the distance within 6 – 12 meters [17-20]. Attempts to suppress the background of fast neutrons by means of sectioning the detector has given the required result. The signal/background ratio has been improved from 0.3 to 0.6. Sectioning of the detector on the criterion of an additional selection for neutrino events has allowed to lower an accidental coincidence background, negatively affecting the accuracy of measurements. Besides, sectioning of the detector makes it possible to keep off watching false effects from the reactor fast neutrons.

For the first time for research reactors difference between experimental antineutrino spectrum and calculated for $^{235}$U spectrum was observed. It was shown, this difference does not depend on distance to reactor center in range 7.3 m – 11.1 m. It allows us to exclude interpretations of spectral distortions as oscillation effect. These distortions of spectrum should be described mostly as a "hole" around 3 MeV than "bump" around 5 MeV (it is principal for searching calculations deviations). After corrections of expected spectrum, we managed to analyze limitations for oscillation to a sterile neutrino and preliminary at 95% CL exclude area of parameters which corresponds to reactor and gallium anomaly. However, accuracy of the experiment should be increased.

## 11. FURTHER PROSPECTS NEUTRINO-4 EXPERIMENT

One of the principal problems, the Neutrino-4 experiment faced in making measurements on the SM-3 reactor, is the correlated background from fast neutrons induced by space radiation. It is caused by the fact that the building with an experimental installation is located on the Earth's surface and the concrete protection above the installation is not sufficiently thick – 1.5 m of concrete only.

Impact of accidental coincidence background on estimating accuracy can be decreased by raising gadolinium concentration in a liquid scintillator. It will allow to reduce the time window for a delayed signal from gadolinium gamma quanta to occur, thus diminishing probability of accidental coincidence. For the correlated background to be suppressed, it is necessary to involve the method of discriminating signals according to an impulse form, which enables to distinguish signals from heavy and light particles. For NEOS (Korea) experiment, one has elaborated and constructed a high-quality scintillator capable of recognizing signals according to an impulse shape [21] which results in considerable lowering the level of the correlated background. Moreover, concentration of gadolinium in this scintillator is 5 times higher than that in our scintillator, which enables to decrease the background influence on an accidental coincidence.

Therefore, the first part of the project involves replacing the currently used scintillator with a new highly efficient one for discriminating signals according to an impulse shape, and with an increased concentration of gadolinium up to 0.5%. It is expected to reduce 3 times the accidental coincidence background as well as the correlated cosmic one, and to increase twice the current measurement precision. Also, active shielding will be improved. Project is planned to be realized with the participation of colleagues from collaborations NEOS [22] and DANSS [23].

The second part of the project is to create a new neutrino laboratory focused on solving essentially important problems outside the framework of the current experiment: monitoring of a cosmic background employing a double detector technique, providing the required energy resolution for a detector by applying the double PMT technique on opposite sides of a scintillator, enhancing the range of measurements up to 14 meters.

In room No. 170, in the SM-3 reactor building, there is a place for a neutrino laboratory, in which it is possible to measure antineutrino flux within the range of 6 – 14 meters. The scheme of the installation is shown in Fig. 22.

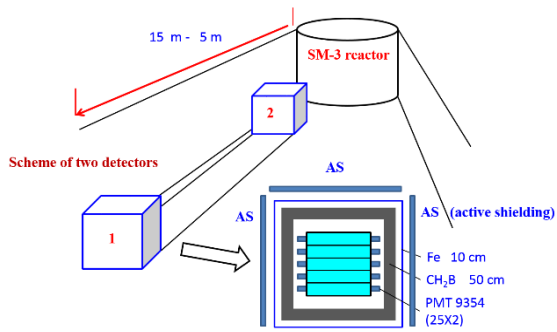

FIG. 22. Scheme of a new experiment on search for neutrino oscillations in room No. 170 of the SM-3 reactor.

In correspondence with a new project, two identical multi-sectional 3D detectors are supposed to be used (50 horizontal sections with PMTs at both ends will set up in orthogonal directions in different planes). One will be installed at maximum distance (14m) for a permanent operation, with the reactor antineutrino flux at this distance being rather weak, while the other will be moving from the first detector to the reactor. Passive and active shielding will be transported simultaneously with the second detector. The first detector is supposed to operate as a monitor, controlling change of the cosmic background. One also regards the option for applying a segmented detector, as mentioned before and shown in Fig. 22. According to preliminary estimates, in two years of collecting data, we expect to obtain statistical accuracy at the level of 1-2% by measuring an antineutrino flux from the reactor. Thus, the problem of possible existence of a sterile neutrino with parameters in the range of $\Delta m_{14}^2 \approx 1 eV^2$ and $\sin^2(2\theta_{14}) < 0.1$ will be clarified. Fig. 23 presents possible restrictions for oscillations parameters which can be achieved in planned experiment.

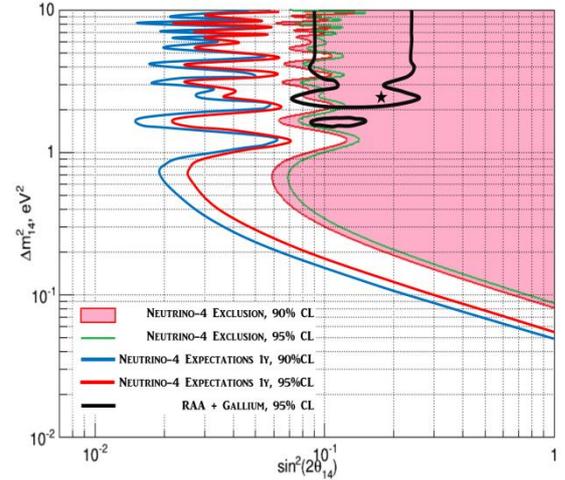

FIG. 23. Restrictions for oscillations parameters which can be achieved in planned experiment.

## 12. RESTRICTIONS FROM OTHER EXPERIMENTS

Fig. 24 illustrates restrictions from other experiments DANSS [23], NEOS [13], STEREO [24] and PROSPECT [25].

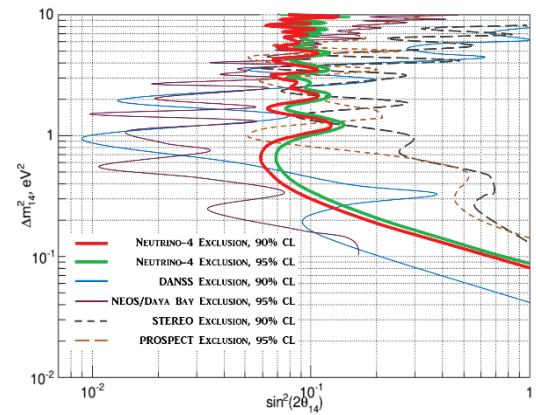

FIG. 24. Comparison of Neutrino-4 restrictions with results of other experiments.

Experiments at nuclear power plants DANSS and NEOS have a quite high sensitivity owing to antineutrino flux higher in order of magnitude than flux at research reactors and at the same time much less background from cosmic rays. Experiment Neutrino-4 has some certain advantages in sensitivity due to wider range of detector movements.

Nevertheless, carrying out measurements at small distances on research reactors is essentially important for a number of reasons. Research reactors operate on the enriched uranium, hence, uncertainty associated with a neutrino source range can be eliminated. The active zone of research reactors, as compared with a nuclear power plant, is not required to consider effects of burning out of fuel. Finally, effects of oscillations have been predicted at the distances of 6 – 14 m, so direct checking of oscillation effects is preferable and model independent. Moreover, an extended model with three sterile neutrinos requires detailed measurements at small distances both research reactors and neutrino sources like experiment BEST at Baksan Neutrino Observatory [26,27].

**ACKNOWLEDGMENTS**

The authors are grateful to the Russian Foundation of Basic Research for support under Contract No. 14-22-03055-ofi_m. The delivery of the scintillator from the laboratory headed by Prof. Jun Cao (Institute of High Energy Physics, Beijing, China) has made a considerable contribution to this research.